# NOTES ON CALCULATING TEMPERATURE OF THE SUPERCONDUCTING TRANSITION: INTERACTING FERMI GAS AND PHONON MECHANISMS IN NON-ADIABATIC REGIME

Lev P. Gor'kov

*NHMFL, Florida State University, 1800 East Paul Dirac Drive, Tallahassee Florida 32310,USA and L.D. Landau Institute for Theoretical Physics of the RAS, Chernogolovka,142432,Russia*

We analyze the mathematical structure of equations for temperature $T_C$ of the superconductivity transition in a gas of interacting Fermi particles or at the phonon-mediated pairing in a metal in case of non-adiabatic conditions $\omega_0 \geq E_F$, i. e., when the characteristic phonon frequency $\omega_0$ is comparable or larger than the Fermi energy $E_F$. The integral equations for $T_C$ are derived in the logarithmical approximation. The equations contain no divergent terms in the anti-adiabatic limit.

## I. INTRODUCTION

The forty-years-old consensus in the theoretical literature concerning stringent constraints on the temperature of transition for phonon mediated superconductivity ( $T_C < 30 \div 40K$ [1]) has recently been challenged by the discovery of superconductivity with the critical temperature as high as $T_c \geq 100K$ in the single layer FeSe deposited on the SrTiO$_3$ substrate [2,3]. The angle resolved photoemission spectroscopy (ARPES) reveals [4] that the band FeSe electrons interact with a surface phonon mode with the frequency $\omega_0 \sim 80 \div 100 meV$. The bottom of the electron pocket at the M-point lies only $60 meV$ below the chemical potential so that the Fermi energy $E_F$ is *smaller* than frequency $\omega_0$ of the active phonon. This is in sharp contrast to ordinary superconductors in which the Debye temperature $\theta_D \sim 100 \div 300K$ is small $\theta_D \ll E_F$.

In metals the extension of the weak-coupling BCS model to the electron-phonon interactions of the arbitrary strength is provided by the set of Migdal-Eliashberg equations [5, 6]. However the applicability of the latter is *subject to the condition of small* adiabatic parameter $\omega_0 / E_F \ll 1$ [5]. In the single layer FeSe on the strontium titanate substrate $\omega_0 / E_F \approx 1.3 \div 1.7$ and this criterion is not fulfilled.

Another case of the violated adiabatic Migdal provision is superconductivity in bulk SrTiO$_3$ in which at low doping the Fermi energy is surprisingly small ($\sim 1 meV$; $T_C \approx (0.07 \div 0.2K)$ [7, 8]). As to the phonons spectrum in SrTiO$_3$, the latter stretches from the acoustic branches to the high frequency optical mode with $\omega_0 \sim 100 meV$ [9]). This case thereby presents the example of the phonon-mediated superconductivity in the *extreme anti-adiabatic* limit $\omega_0 \gg E_F$.

Note that with varying the ratio between $\omega_0$ and $E_F$ also somewhat changes the very concept of the phonon pairing. From condition $\omega_0/E_F \ll 1$ in the adiabatic regime follows the celebrated retardation effect of the BCS theory. Indeed, each of the two electrons comprising the Cooper pair for the characteristic phonon time $1/\omega_0$ shifts by $d \sim v_F/\omega_0 \sim (1/p_F)(E_F/\omega_0)$. Thereby the effect of the Coulomb repulsion is reduced because the latter is screened on the atomic scale; meanwhile the electrons of the pair stay apart from each other on distances $d$ larger than the typical interatomic distance $a \approx 1/p_F$ by the factor $E_F/\omega_0 \gg 1$.

In the opposite limit $\omega_0/E_F \gg 1$ electrons of the pair sense instantaneously and at the same time both the direct Coulomb repulsion and the potential created by the local lattice distortion. Therefore for the Cooper pairs to form the strength of the phonon attraction must *prevail* over the direct electron-electron Coulomb repulsive interaction.

At $\omega_0 \sim E_F$ the electron-electron potential interaction must be treated on an equal basis with attraction via the virtual exchange by a phonon.

Changes in the underlying physical picture must find reflection in the *mathematical apparatus* of the theory. In conventional metals contributing to the superconducting gap are only the electrons within a narrow vicinity of the Fermi energy $\Delta\varepsilon \approx \omega_0 \ll E_F$. Convergence of the logarithmic integrals in the Cooper channel in this case is guaranteed by the phonon Green function [6, 10]. However, in the opposed limit $\omega_0 \gg E_F$ the question arises what replaces the phonon frequency as a cut-off parameter.

The problem is pertinent, in particular, to the physics of cold gases and for the first time was studied for the gas of the neutral Fermi particles with weak attractive interaction [11]. In that follows we extend the method [11] to the case of the non-adiabatic phonon-mediated superconductivity pairing.

The validity of the Migdal-Eliashberg equations [5, 6] has been repeatedly questioned in the theoretical literature on various mechanisms alternative to the phonon-mediated pairing such as the so called plasmon and excitonic mechanisms expected to provide higher $T_C$, or at discussions in regard to the nature of high $T_C$ superconductivity in cuprates. The methodic difficulty inevitably arising is that with $\omega_0/E_F \geq 1$ the key advantage of the Migdal theory, namely, the possibility of neglecting all contributions from the so called "crossing" diagrams is lost. In theoretical publications on the theme the authors are often preoccupied analyzing contributions from particular classes of "crossing" diagrams such, for instance, as the so called vertex corrections [12].

In our opinion, in the absence of small parameters such incremental improvements lead nowhere. Instead, we suggest the analysis in frameworks of the weak-coupling approximation. We argue that such analysis allows studying every qualitative feature pertinent to a particular non-adiabatic mechanism. At the time, the weak coupling BCS theory turned out extremely successful at interpretation of the experimental data.

## II. THE WEAK-COUPLING LIMIT

The canonic BCS weak-coupling expression for temperature $T_c$ of the superconductivity transition has the form:

$$T_c = W \exp(-1/\lambda) \quad (1)$$

Here in (1) $W$ is the order of magnitude cut-off parameter in the Cooper channel. At $\omega_0 \ll E_F$ the phonon-mediated attraction is effective in the narrow vicinity of the energy Fermi. So $W \approx \omega_0$, where $\omega_0$ is the Debye temperature or a typical phonon frequency. (Proportionality $\omega_0 \propto M^{-1/2}$ where $M$ is an atomic mass lies at the core of the isotope effect). In Eq. (1) $\lambda$ is proportional to the product of the interaction constant and the density of states at the Fermi level ($\lambda \ll 1$).

In the limit $\omega_0 \gg E_F$ the only cut-off in the Cooper channel can be the Fermi energy itself. One of the consequences is that such discriminating signature of a phonon mechanisms as the isotope effect is absent in this limit.

Below the integral equations for temperature of the superconductivity transition are analyzed in the general case of $\omega_0 \sim E_F$.

## III. PHONON-MEDIATED SUPERCONDUCTIVITY AT $\omega_0 \sim E_F$ IN GENERAL

As mentioned, the key element of the Migdal [5] theory of the electron- phonon interaction in metals is the possibility to discard in the diagrammatic expansion contributions coming from the so called "crossing diagrams". This greatly advantageous feature at $\omega_0 \sim E_F$ is lost already in the normal phase. Performing the analytic calculations reduces itself to analysis of the perturbation expansion. Correspondingly, below all results related to the Cooper instability in non-adiabatic conditions are obtained in the so called logarithmic approximation.

In particular, in the limit $\omega_0 \gg E_F$ the expression $T_c = const \times E_F \exp(-1/\lambda)$ contains a numerical factor $const \sim 1$. Formally, for validity of the result must be "large" $\ln(E_F/T_c) \gg 1$. In practice, due to the exponential dependence in the expression (1) for the approach to work it is enough for $\ln(E_F/T_c) > 1$. In each concrete case this can be verified if the Fermi energy $E_F$ is also known experimentally.

Signature of the superconducting instability is the pole appearing in the scattering amplitude for two electrons $\Gamma(p, q-p \mid p', q-p')$ at the transition temperature $T_C$ [10]. The amplitude is the sum of all diagrams in the Cooper channel. Commonly one analyzes the diagrammatic series at zero summary momentum and frequency $q = 0$. The resulting equation in the notations $\Gamma(p, q-p \mid p', q-p') \equiv \Gamma(p \mid p')$ reads:

$$\Gamma(p \mid p') = \tilde{\Gamma}(p \mid p') - \frac{T}{(2\pi)^3} \sum_{n'} \int d\vec{k}\, \tilde{\Gamma}(p \mid k) G(k) G(-k) \Gamma(k \mid p') \ . \quad (2)$$

In Eq. (2) $\tilde{\Gamma}(p \mid p')$ represents the block of the so called irreducible diagrams, i.e. diagrams that cannot be cut into the two parts by crossing two parallel electronic lines. From the mathematic viewpoint the very instability in the Cooper channel owes its origin to the logarithmical singularity and divergent contributions at the summation and integrations inside the blocks represented in Eq. (3) by the product of two Green functions $G(k)G(-k)$. The electrons Green function in the thermodynamic technique has the form [10]:

$$G(k) = [iv_n - (\vec{k}^2 - p_F^2)/2m]^{-1}. \quad (3)$$

Generally, the matrix element for electron-electron scattering can be taken in the following form:

$$M(p|p') = V(\vec{p} - \vec{p}') - \gamma^2(\vec{p} - \vec{p}') \times D_0(p - p'). \quad (4)$$

Here $D_0(p - p')$ is the phonons Green function:

$$D_0(p - p') = -\omega_0^2(\vec{p} - \vec{p}')/[(\varepsilon_n - \varepsilon_m)^2 + \omega_0^2(\vec{p} - \vec{p}')], \quad (5)$$

$V(\vec{p} - \vec{p}')$ is the direct Coulomb term; $\gamma(\vec{p} - \vec{p}')$ is the electron-phonon coupling constant. $M(p|p')$ is the first term in the perturbation expansion for the block $\tilde{\Gamma}(p|p')$ in Eq. (2).

Because of its singular character role of the Coulomb interaction in the theory of superconductivity, however, needs special consideration (see [13]). Below we focus on treatment of the interactions mediated by phonons in non-adiabatic conditions. Correspondingly, the Coulomb term $V(\vec{p} - \vec{p}')$ in Eq. (4) will be omitted. From now on:

$$M(p|p') = -\gamma^2(\vec{p} - \vec{p}') \times D_0(p - p'). \quad (4a)$$

## IV. POSING THE QUESTION

To make the transformations below more transparent, the equations are simplified by omitting temporarily dispersion in the spectrum of phonons. In (4a, 5) $\omega_0(\vec{p} - \vec{p}') \equiv \omega_0$:

$$M(p|p') \Rightarrow M(\varepsilon_n | \varepsilon_m) = -\gamma^2 \omega_0^2 / [(\varepsilon_n - \varepsilon_m)^2 + \omega_0^2], \quad (6)$$

and $\Gamma(p|p')$ depends only on the energy variables. Changing the notations again into $\Gamma(p|p') \Rightarrow \Gamma(\varepsilon_n | \varepsilon_m)$ one writes:

$$\Gamma(\varepsilon_n | \varepsilon_{n'}) = \tilde{\Gamma}(\varepsilon_n | \varepsilon_{n'}) - T \sum_m \tilde{\Gamma}(\varepsilon_n | \varepsilon_m) \Pi^{(d)}(\varepsilon_m) \Gamma(\varepsilon_m | \varepsilon_{n'}). \quad (7)$$

The notation $\Pi^{(d)}(\varepsilon_m)$ in Eq. (7) stands for the integral:

$$\Pi^{(d)}(\varepsilon_m) = \int \frac{d\vec{k}}{(2\pi)^d} G(k)G(-k) = \int \frac{d\vec{k}}{(2\pi)^d} \frac{1}{v_m^2 + [(\vec{k}^2 - p_F^2)/2m]^2} . \quad (8)$$

(In (7, 8) and in the equations below index $d$ signifies dimensionality of the problem: $d = 2, 3$).

$M(\varepsilon_n | \varepsilon_m)$ (6) and, hence $\Gamma(\varepsilon_n | \varepsilon_m)$ decrease at large $\varepsilon_n > \omega_0$; therefore *formally*, in Eq. (7) the summation over $\varepsilon_m$ converges at $\varepsilon_m$ of the order of $\omega_0$, the phonon frequency. In reality, such a cut-off makes sense only in a common metal where $\omega_0 \ll E_F$ and both the integration and the summation in (2, 7) are limited to the narrow vicinity of the Fermi energy; in that case one returns to the familiar BCS prefactor $W \sim \omega_0$ in the expression (1) for $T_c$.

In fact, in the opposite limit of $\omega_0 \gg E_F$, one concludes from Eqs. (6, 7) that at $\varepsilon_n < \omega_0$ $\Gamma(\varepsilon_n | \varepsilon_{n'})$ is a constant $\Gamma(\varepsilon_n | \varepsilon_{n'}) \equiv \bar{\Gamma}$. However, substituting constant $\Gamma(\varepsilon_n | \varepsilon_{n'})$ into the right hand side of (7) one obtains the diverging expression.

The difficulty has the obvious origin. At high energies, that is, at $\vec{k}^2 \gg p_F^2$ electrons in the Fermi gas are indistinguishable from free electrons. Far from the Fermi surface the product $G(k)G(-k)$ goes over into the product $G^{(0)}(k)G^{(0)}(-k)$ of the two Green functions for the two Fermi particles *in vacuum*:

$$G^{(0)}(k)G^{(0)}(-k) = \frac{1}{\varepsilon_m^2 + [\vec{k}^2 / 2m]^2} \quad . \quad (9)$$

That is, the contribution from $\Pi^{(d)}(\varepsilon_m)$ (8) at large $\varepsilon_m$ coincides with the expression for the second-order Born correction to the scattering amplitude for two *free* electrons and the divergence in (7) must be removed by properly renormalizing the interaction [11].

As shown below, integrations in (7) actually converge at $\varepsilon_m \sim \mu$ and in the anti-adiabatic regime role of the cut-off parameter in Eq. (1) belongs to the Fermi energy itself $W \sim E_F$.

## V. LOGARITHMIC APPROXIMATION IN THE GENERAL CASE

In general terms, the "logarithmic accuracy" signifies the approximation where the smallness of the product $\nu(E_F)M(p|p') \ll 1$ of the matrix element and the density of states is compensated by a large logarithmic factor $\nu(E_F)M(p|p') \times \ln(W/T) \sim 1$.

Return to the general case of $\omega_0 \sim E_F$ and rewrite Eq. (2) in the form ready for use both in three and two dimensions ($d=2, 3$):

$$\Gamma(p | p') = \tilde{\Gamma}(p | p') - \frac{T}{(2\pi)^d} \sum_{n'} \int d\vec{k} \tilde{\Gamma}(p | k) G(k) G(-k) \Gamma(k | p') \, . \, (2')$$

Restoring dispersion in the phonons spectrum in Eq. (4a), for the bare vertex $\Gamma^{(1)}(p | p')$ in all equations we assume the expression $\Gamma^{(1)}(p | p') \equiv M(p | p')$ from Eq. (4a).

Define the scattering amplitude $\Gamma^{(0)}(p | p')$ for two particles *in vacuum* by the equation:

$$\Gamma^{(0)}(p\,|\,p') = \Gamma^{(1)}(p\,|\,p') - \frac{T}{(2\pi)^d}\sum_{n'}\int d\vec{k}\,\Gamma^{(1)}(p\,|\,k)G^{(0)}(k)G^{(0)}(-k)\Gamma^{(0)}(k\,|\,p') \quad (10)$$

Inverting Eq. (10) introduces the operator $\hat{L}$:

$$\hat{L}\Gamma^{(0)}(p\,|\,p') \equiv \Gamma^{(0)}(p\,|\,p') + \frac{T}{(2\pi)^d}\sum_{n'}\int d\vec{k}\,\Gamma^{(1)}(p\,|\,k)G^{(0)}(k)G^{(0)}(-k)\Gamma^{(0)}(k\,|\,p'). \quad (11)$$

Rewrite Eq. (2') in the form:

$$\Gamma(p\,|\,p') + \frac{T}{(2\pi)^d}\sum_{n'}\int d\vec{k}\,\Gamma^{(1)}(p\,|\,k)G^{(0)}(k)G^{(0)}(-k)\Gamma^{(0)}(k\,|\,p')$$
$$= \Gamma^{(1)}(p\,|\,p') + \Gamma^{(2)}(p\,|\,p') - \frac{T}{(2\pi)^d}\sum_{n'}\int d\vec{k}\,\Gamma^{(1)}(p\,|\,k)[G(k)G(-k) - G^{(0)}(k)G^{(0)}(-k)]\Gamma(k\,|\,p')$$
$$- \frac{T}{(2\pi)^d}\sum_{n'}\int d\vec{k}\,\Gamma^{(2)}(p\,|\,k)G(k)G(-k)\Gamma(k\,|\,p'). \quad (12)$$

In Eq. (12) $\Gamma^{(1)}(p\,|\,p') \equiv M(p\,|\,p')$ and $\Gamma^{(2)}(p\,|\,p')$ are the irreducible diagrams of the second and higher order in $M(p\,|\,p')$.

From (10, 11) follows:

$$\hat{L}\Gamma(p\,|\,p') = \hat{L}\Gamma^{(0)}(p\,|\,p') + \Gamma^{(2)}(p\,|\,p')$$
$$- \frac{T}{(2\pi)^d}\sum_{n'}\int d\vec{k}\,\hat{L}\Gamma^{(0)}(p\,|\,k)[G(k)G(-k) - G^{(0)}(k)G^{(0)}(-k)] \quad (13)$$
$$- \frac{T}{(2\pi)^d}\sum_{n'}\int d\vec{k}\,\Gamma^{(2)}(p\,|\,k)G(k)G(-k)\Gamma(k\,|\,p')$$

Applying the inverse operator $\hat{L}^{-1}$ to the both sides of Eq. (13) one finally arrives to the following equation:

$$\Gamma(p\,|\,p') = \Gamma^{(0)}(p\,|\,p') + \Gamma^{(2)}(p\,|\,p') - \frac{T}{(2\pi)^d}\sum_{m}\int d\vec{k}\,\Gamma^{(0)}(p\,|\,k)\tilde{\Pi}^{(d)}(k)\Gamma(k\,|\,p')$$
$$- \frac{T}{(2\pi)^d}\sum_{n'}\int d\vec{k}\,\Gamma^{(2)}(p\,|\,k)G(k)G(-k)\Gamma(k\,|\,p'). \quad (14)$$

(Restricting by the first Born approximation in Eq. (10) for the Fermi particles *in vacuum* one has $\Gamma^{(0)}(p\,|\,p') \simeq M(p\,|\,p')$; $\hat{L}^{-1}\Gamma^{(2)}(p\,|\,p') \approx \Gamma^{(2)}(p\,|\,p')$). In the upper line of Eq. (14) $\tilde{\Pi}^{(d)}(k)$ is the difference:

$$\tilde{\Pi}^{(d)}(k) = [G(k)G(-k) - G^{(0)}(k)G^{(0)}(-k)]. \quad (15)$$

Convergence at the summation and integration over the momentum in the upper line of Eq. (14) is now guaranteed both by the presence of the phonons Green function in Eq. (4) and by the fact that the block $\tilde{\Pi}^{(d)}(k)$ in (15) decreases at large $k = (\varepsilon_m, \vec{k})$.

To determine temperature of the superconductivity onset one must find the *eigenvalue* of the homogeneous equation (14). Rewriting $\Gamma(p|p') \Rightarrow \psi(p) \times \psi(p')$, from (14) follows the equation for the function $\psi(p)$:

$$\psi(p) = -\frac{T}{(2\pi)^d}\sum_m \int d\vec{k} M(p|p')\tilde{\Pi}^{(d)}(k)\psi(k) - \frac{T}{(2\pi)^d}\sum_{n'}\int d\vec{k}\Gamma^{(2)}(p|k)G(k)G(-k)\psi(k). \quad (14a)$$

## VI. MATRIX EQUATION IN CASE OF DISPERSIONLESS PHONONS

It is instructive to solve Eq. (14a) in the case of $M(\varepsilon_n|\varepsilon_m) \equiv -\gamma^2\omega_0^2/[(\varepsilon_n - \varepsilon_m)^2 + \omega_0^2]$, Eq. (6). In this example most calculations can be performed to the end.

Denote solution of the homogeneous Eq. (14) for disperssionless phonons as $\psi(k) \Rightarrow \psi(\varepsilon_n)$. One has (compare with Eq. (14a)):

$$\psi(\varepsilon_n) = -T\sum_m M(\varepsilon_n|\varepsilon_m)\tilde{\Pi}^{(d)}(\varepsilon_m)\psi(\varepsilon_m) - \frac{T}{(2\pi)^d}\sum_{n'}\int d\vec{k}\Gamma^{(2)}(p|k)G(k)G(-k)\psi(\varepsilon_m) \quad (16)$$

(For derivation of the expressions for the kernel $\tilde{\Pi}^{(d)}(\varepsilon_m)$ at $d = 2, 3$, see in Appendix I). In two dimensions:

$$\tilde{\Pi}^{(2)}(\varepsilon_n) = \left(\frac{m}{2\pi}\right)\left(\frac{1}{\varepsilon_n}arctg\frac{\mu}{\varepsilon_n}\right). \quad (17a)$$

For three dimensions:

$$\tilde{\Pi}^{(3)}(\varepsilon_n) = \frac{(2m)^{3/2}}{2\pi}\left(\frac{1}{2\varepsilon_n}\right)\left[\sqrt{\mu + i\varepsilon_n} + \sqrt{\mu - i\varepsilon_n} - \sqrt{i\varepsilon_n} - \sqrt{-i\varepsilon_n}\right]. \quad (17b)$$

($\mu \equiv E_F$). Without terms quadratic in $M(\varepsilon_n|\varepsilon_m)$ Eq. (15) acquires the transparent form of the matrix equation:

$$\psi(\varepsilon_n) = -T\sum_m M(\varepsilon_n|\varepsilon_m)\tilde{\Pi}^{(d)}(\varepsilon_m)\psi(\varepsilon_n). \quad (18)$$

To emphasize once again, $\tilde{\Pi}^{(d)}(\varepsilon_m)$ in the expressions (17a, b) decreases at $\varepsilon_m > E_F$, thereby convergence of the summation in (18) is guaranteed at *any ratio* between $\omega_0$ and $E_F$.

At the arbitrary $\omega_0$ and $E_F$, Eq. (18) will be solved numerically elsewhere. In the limit $\omega_0 \ll E_F$ one returns to the BCS result.

In the opposite limit $\omega_0 \gg E_F = \mu$ the main contribution in (18) comes from $\varepsilon_m \ll \omega_0$. At $\varepsilon_n, \varepsilon_m \ll \omega_0$ $M(\varepsilon_n | \varepsilon_m)$ is a constant ($M(\varepsilon_n | \varepsilon_m) \simeq -\gamma^2$) and in the anti-adiabatic limit the temperature $T_c$ is defined by the algebraic equation:

$$1 = \gamma^2 T \sum_m \tilde{\Pi}(\varepsilon_m). \quad (19)$$

(Summation over $\varepsilon_m$ in $T \sum_m \tilde{\Pi}(\varepsilon_m)$ is carried out in Appendix II).

Before proceeding further, return however to the contribution from the last term in Eqs. (14,14a and 16) containing $\Gamma^{(2)}(p|k)$. $\Gamma^{(2)}(p|k)$ is of the second order in $M(p|p')$. One can verify that all second order diagrams constituting $\Gamma^{(2)}(p|k)$ belong to the class of "crossing" diagrams [11]. Therefore in the adiabatic limit $\omega_0 \ll E_F$ this term is negligibly small.

At $\omega_0 \gg E_F$ the logarithmic contribution that comes about from summation and the integrations of the product of the two Green functions $G(k)G(-k)$ in (16) is multiplied by a factor that must be calculated by performing the internal integrations in the diagrammatic expression for $\Gamma^{(2)}(p|k)$. Since $\Gamma^{(2)}(p|k)$ is *quadratic* in $M(p|p')$, the result contributes only into the numeric coefficient in front of the expression for $T_c$. After simple but somewhat tedious calculations one obtains:

$$1 = \gamma^2 \frac{m}{\pi \hbar^2} \ln\left(\frac{2\mu\gamma}{\pi e^2 T}\right) \quad (20a)$$

in 2D and in the 3D case [11]:

$$1 = \gamma^2 \frac{m p_F}{2\pi^2 \hbar^3} \ln\left[\left(\frac{2}{e}\right)^{7/3} \frac{\mu\gamma}{\pi T}\right]. \quad (20b)$$

Substituting $\mu \equiv E_F$, for the temperature of transition follows:

$$T_C^{(2d)} = \left(\frac{2\gamma\mu}{\pi e^2}\right) \times \exp\left(-\frac{\pi \hbar^2}{m\gamma^2}\right) \simeq 0.15 E_F \times \exp\left(-\frac{\pi \hbar^2}{m\gamma^2}\right) \quad (21a)$$

and

$$T_C^{(3D)} = \frac{\gamma}{\pi}(\frac{2}{e})^{7/3} E_F \exp\left[-\frac{2\pi^2 \hbar^3}{m p_F \gamma^2}\right] \approx 0.27 E_F \exp\left[-\frac{2\pi^2 \hbar^3}{m p_F \gamma^2}\right] \quad (21b)$$

Depending on value of the parameters in the exponent and the value of $E_F$ itself, $T_c$ may be higher than one can expect in the adiabatic limit.

### VII. TEMPERATURE OF TRANSITION WITH LOGARITHMIC ACCURACY

In its new form the integral equation (14) is encumbered by the presence of the additional contribution with $\Gamma^{(2)}(p|k)$. Singling out the logarithmic factor from the block of $G(k)G(-k)$, this contribution is proportional to $\propto \Gamma^{(2)}(p|0)\nu(E_F) \times \ln(W/T)\Gamma(0|p')$ (here $(\ldots|0)$ signifies the choice of $p$ or $k$ at the Fermi level). Eq. (14a) acquires the form:

$$\psi(p) = -\frac{T}{(2\pi)^d}\sum_m \int d\vec{k}\,\Gamma^{(0)}(p|k)\tilde{\Pi}^{(d)}(k)\psi(k) + F(p)\psi(0) \,. \quad (14b)$$

In principle, both $\Gamma^{(2)}(p|0) \propto M^2$ (and $F(p) \propto M$) can be calculated at the arbitrary form of interactions in (4, 5), but calculations are tedious and the resulting expressions are not useful.

Meanwhile, in the limiting case of a constant $\Gamma^{(1)}(\varepsilon_n|\varepsilon_{n'}) \approx -\gamma^2$ *the exact solution* was obtained in Sect.VI and it was shown that taking the $\Gamma^{(2)}(p|k)$-terms into account changes only the *numeric factor* of the order of unity in the expression (1) for $T_C$. With such accuracy the homogeneous weak-coupling Eq. (14) can be solved *directly*.

In fact, in Eq. (2) write:

$$-\frac{T}{(2\pi)^3}\sum_{n'} \int d\vec{k}\,\tilde{\Gamma}(p|k)\tilde{\Pi}^{(d)}(k)\Gamma(k|p')$$

$$\Rightarrow -T\sum_{n'} \int \tilde{\Gamma}(p|k)[mp_F \sin\theta d\theta]/(2\pi)^2]d\varsigma\,\tilde{\Pi}^{(d)}(k)\Gamma(k|p') \,. \quad (22)$$

(In (22) $\varsigma = (\vec{k}^2 - p_F^2)/2m \approx v_F(p - p_F)$; $\theta$ is the angle between two vectors $\vec{p}$ and $\vec{k}$. $\tilde{\Pi}^{(d)}(k)$ is from Eq. (15)).

Let $\varepsilon_n = \varepsilon_m$ and the vectors $\vec{p}$ and $\vec{k}$ be *on the Fermi surface*. The expression for $\tilde{\Gamma}(p|k) \equiv \tilde{\Gamma}(\theta)|_{FS}$ defines the factor in front of the logarithmic singularity in Eq. (22):

$$\left[\int_0^\pi \sin\theta d\theta\,\tilde{\Gamma}(\theta)|_{FS}\right]\frac{mp_F}{(2\pi)^2} \times \int_0^W \frac{d\varsigma}{\varsigma}\,\text{th}\frac{\varsigma}{2T} \Rightarrow \lambda \ln\left(\frac{2W\gamma}{\pi T}\right). \quad (23)$$

Eq. (23) provides definition of $\lambda$ in $T_C = \text{const} W \exp(-1/\lambda)$.

For the problem in hand $\tilde{\Gamma}(p|k)|_{FS} = -\gamma^2(2p_F(1-\cos\theta))$. As to the prefactor, $W \approx \omega_0$, in the adiabatic limit, and $W \approx E_F$ in the extreme case of $\omega_0 \gg E_F$. In details the variation of the cut-off parameter $W$ from $W \approx \omega_0$ to $W \approx E_F$ can be investigated (omitting the last term with $\Gamma^{(2)}(p|k)$) from the homogeneous Eq. (14):

$$\psi(p) = -\frac{T}{(2\pi)^d}\sum_m \int d\vec{k}\,\Gamma^{(0)}(p|k)\tilde{\Pi}^{(d)}(k)\psi(k), \quad (24)$$

where $\Gamma^{(0)}(p|k) = M(p|k)$ from Eq. (4a).

## VIII. CONCLUSION AND SUMMARY

The Migdal theory of electron-phonon interaction is not applicable in the non-adiabatic regime and the consistent analysis of equations for temperature of the superconductivity transition is possible only in the weak-coupling approximation. We concentrated on studying the Cooper instability for the phonon-mediated attraction in general case.

It was shown that at extension of the BCS-like formulation on a non-adiabatic regime in the diagrammatic Eqs. (2, 2') appear unphysical contributions that must be removed by subtracting the second order Born corrections to the scattering amplitudes for the Fermi particles *in vacuum*.

In the new form the equations for temperature $T_c$ of the superconductivity transition are given by the integral equations (14a, b).

In addition to several exact results obtained in the anti-adiabatic limit for dispersionless phonons, the transition temperature $T_c$ was found with the logarithmic accuracy in general case and at any value of the Migdal parameter $\omega_0 / E_F$.

The method can be applied to arbitrary short-range interactions between Fermi particles and for the anisotropic band spectrum in metals.

## ACKNOWLEDGMENTS

The work was supported by the National High Magnetic Field Laboratory through NSF Grant No. DMR-1157490, the State of Florida and the U.S. Department of Energy.

## APPENDIX I

For the isotropic spectrum the integration over momentum $\vec{k}$ in

$$\tilde{\Pi}^{(d)}(\varepsilon_m) = \int d\vec{k}(2\pi)^{-d}[G(k)G(-k) - G^{(0)}(k)G^{(0)}(-k)]$$

is performed analytically by integrating over the variable $u = 2mk^2$. In 2D calculation of $\tilde{\Pi}^{(2)}(\varepsilon_n)$ in Eq.(15) reduces to the standard integrals.

In 3D the expression (17b) for $\tilde{\Pi}^{(3)}(\varepsilon_n)$ is presented in the form:

$$\tilde{\Pi}^{(3)}(\varepsilon_n) = \frac{(2m)^{3/2}}{(2\pi)^2}\int_o^\infty u^{1/2}du\left\{\frac{1}{\varepsilon_n^2 + (u-\mu)^2} - \frac{1}{\varepsilon_n^2 + u^2}\right\}$$

Rotating the *contour* $\{0;\infty\}$ in the complex plane by $2\pi$ and calculating the residues one arrives to the expression Eq. (17b).

## APPENDIX II

Transform the sum over $\varepsilon_m$ in $T\sum_m \tilde{\Pi}(\varepsilon_m)$ into the integrals; in 2D

$$T\sum_m \tilde{\Pi}^{(2)}(\varepsilon_m) = \frac{1}{2}\int_0^\infty du\left\{\frac{th(u-\mu)/2T}{u-\mu} - \frac{th(u/2T)}{u}\right\}$$

and in 3D

$$T\sum_m \tilde{\Pi}^{(3)}(\varepsilon_m) = \frac{1}{2}\int_0^\infty u^{1/2}du\left\{\frac{th(u-\mu)/2T}{u-\mu} - \frac{th(u/2T)}{u}\right\}$$

After integrating by parts and making use of relation $T_c \ll \mu$ one arrives to the results in Eqs. (20, 21) of the main text.